\documentclass[aps,prl,reprint,superscriptaddress]{revtex4-1}

\usepackage{color}
\usepackage{soul}
\usepackage{dcolumn}
\usepackage{graphicx}
\usepackage{amsmath}
\usepackage{amssymb}
\usepackage{bm}
\usepackage{booktabs}
\begin{document}
\mbox{}

\title{Tuning the inductance of Josephson junction arrays without SQUIDs}

\author{R. Kuzmin}
\affiliation{Department of Physics, University of Maryland, College Park, Maryland 20742, USA.}
\author{N. Mehta}
\affiliation{Department of Physics, University of Maryland, College Park, Maryland 20742, USA.}
\author{N. Grabon}
\affiliation{Laboratory for Physical Sciences, College Park, Maryland 20740, USA.}
\author{V. E. Manucharyan}
\affiliation{Department of Physics, University of Maryland, College Park, Maryland 20742, USA.}
\affiliation{École Polytechnique Fédérale de Lausanne (EPFL), Lausanne CH-1015, Switzerland.}

\date{\today}
\begin{abstract}
It is customary to use arrays of superconducting quantum interference devices (SQUIDs) for implementing magnetic field-tunable inductors. Here, we demonstrate an equivalent tunability in a (SQUID-free) array of single Al/AlOx/Al Josephson tunnel junctions. With the proper choice of junction geometry, a perpendicularly applied magnetic field bends along the plane of the superconductor and focuses into the tunnel barrier region due to a demagnetization effect. 
Consequently, the Josephson inductance can be efficiently modulated by the Fraunhoffer-type supercurrent interference. The elimination of SQUIDs not only simplifies the device design and fabrication, but also facilitates a denser packing of junctions and, hence, a higher inductance per unit length. As an example, we demonstrate a transmission line, the wave impedance of which is field-tuned in the range of $4-8~\textrm{k}\Omega$, centered around the important value of the resistance quantum $h/(2e)^2 \approx 6.5~\textrm{k}\Omega$.
\end{abstract}

\maketitle

The arrays of Josephson junctions are widely used in science and technology as compact on-chip low-loss kinetic inductors operating well into the microwave frequency range. Thanks to the Josephson effect, the kinetic inductance density in such arrays can exceed the vacuum permeability by four orders of magnitude \cite{Manucharyan2012Superinductance}. This property enables access to new regimes of quantum fluctuations in superconducting circuits \cite{Watanabe2001}, demonstrations of novel superconducting qubits \cite{Manucharyan2009,Kalashnikov2020,Pechenezhskiy2020,Gyenis2021}, as well as applications in parametric amplification \cite{Nation2012,Castellanos-Beltran2008,Macklin2015,Krupko2018,Planat2020},
electromechanical transduction \cite{Arrangoiz-Arriola2018}, 
and hybrid circuit QED \cite{Stockklauser2017}. Of particular interest is the use of Josephson arrays for creating electromagnetic vacuums with a characteristic impedance $Z$ exceeding the scale of the resistance quantum $R_Q=h/(2e)^2\approx6.5~\textrm{k}\Omega$, which enable new regimes of quantum electrodynamics \cite{Kuzmin2019npj}, simulations of quantum impurity problems \cite{Kuzmin2021,Leger2022,Goldstein2013,Houzet2020,Burshtein2021}, explorations of the superconductor-insulator quantum phase transitions \cite{Chow1998,Miyazaki2002,Bard2017,Cedergren2017,Kuzmin2019sit}, and possibly a metrology of dc current \cite{Kuzmin1991,Weissl2015,Shimada2016,Wang2019,Shaikhaidarov2022,Crescini2022}. Finally, Josephson junction arrays are building blocks for various wire-based Josephson metamaterials \cite{Jung2014} with applications in tunable dielectrics \cite{Trepanier2019} and dark matter detectors \cite{Gelmini2020}.

\begin{figure}[htbp]
	\centering
	\includegraphics[width=1\linewidth]{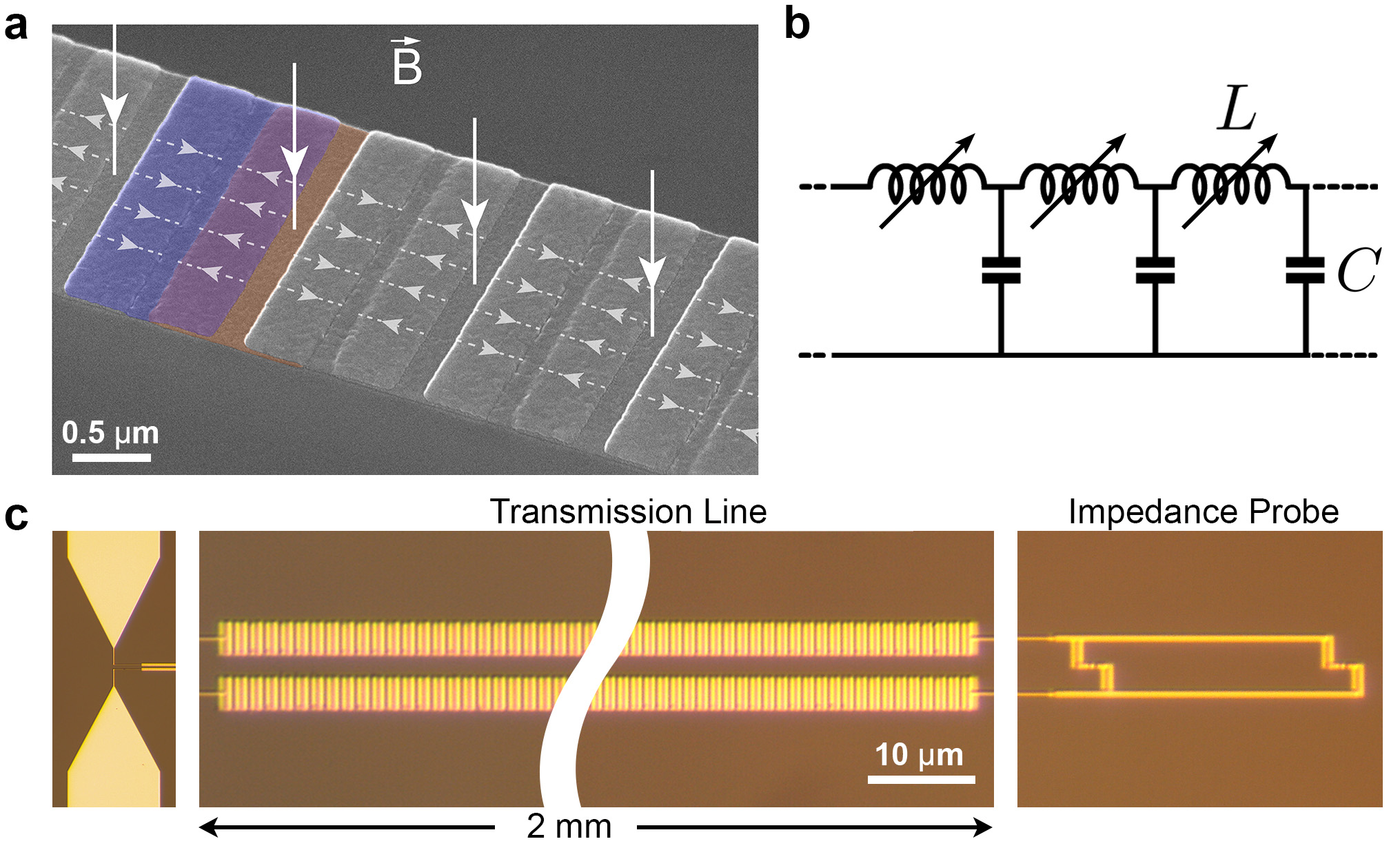}
	\caption{(a) Scanning electron microscope image of a section of a Josephson junction array. The junctions are formed by an overlap of aluminum islands, with an example shown in false colors (blue-top, orange-bottom), separated by a thin barrier of aluminum oxide (not shown). Only two islands are highlighted. Dashed arrows illustrate the magnetic field lines penetrating through the barrier layers when the array is placed in a transverse magnetic field (solid white arrows).
	(b) A circuit diagram of a Josephson junction array well-described in a long-wavelength limit as a telegraph transmission line. (c) An optical photograph of a tunable Josephson transmission line comprised of two parallel junction arrays (at the center) connected to a dipole antenna on the left and to a qubit on the right. The qubit plays the role of a probe for the transmission line's impedance.
	}

	\label{fig:Fig1}
\end{figure}

Tuning the array's inductance in-situ is a very useful feature for many of the applications mentioned above, see e.g. \cite{Watanabe2001control,PuertasMartinez2019,Leger2019,Rastelli2018}. To date, a common way to tune the array's inductance, is to split each junction into two, forming a SQUID array, and piercing the SQUID loops with a global external magnetic field. In this work, we demonstrate an overlooked method to flux-tune an array's inductance without introducing SQUIDs. Our scheme relies on a demagnetization effect in standard rectangular overlap-type Josephson junctions placed in a transverse magnetic filed. As was noticed quite some time ago, a magnetic field perpendicular to a junction's barrier creates demagnetizing currents in the junction's electrodes \cite{Rosenstein1975,Hebard1975}. These currents produce a magnetic field which penetrates the barrier region (Fig. 1a) and modulates the junction's critical current $I_c$. Later experiments and simulations demonstrated that the effect of the transverse magnetic field on $I_c$ depends strongly on the junction geometry \cite{Monaco2008,Yeh2012}. In fact, with short and wide junctions, like in Fig. 1a, the perpendicular field can be much more capable in the modulation of $I_c$ than the in-plane one \cite{Monaco2009}. We use this effect to simultaneously tune the inductance of many thousands of junctions in a Josephson junction array.

To fully appreciate the benefits of our approach, let us recall that every Josephson junction array comes with some stray capacitance and behaves in a long-wavelength limit like a telegraph transmission line (Fig. 1b) \cite{Masluk2012}. The array's inductance $L$ and capacitance $C$ per unit cell define the wave velocity $v=1/\sqrt{LC}$ and impedance $Z=\sqrt{L/C}$. Thanks to the small size of a single Josephson junction, their arrays can comprise tens of thousands of cells and form Josephson transmission lines with $Z\gg R_Q$ while staying low-loss \cite{Kuzmin2019sit}. Because of the extra metal in the SQUID loop, the $C$ in a SQUID array is larger than in an array of single junctions with the same junction's area. This makes it harder for SQUID arrays to reach the range of $Z\gtrsim R_Q$, which is required in many applications. Another challenge with SQUIDs is making an array of a large length, as SQUIDs are greater in size than single junctions. Finally, the arrays of single junctions are simpler structures, and therefore they should be more prone to disorder than the SQUID arrays.

\begin{figure}
	\centering
	\includegraphics[width=1\linewidth]{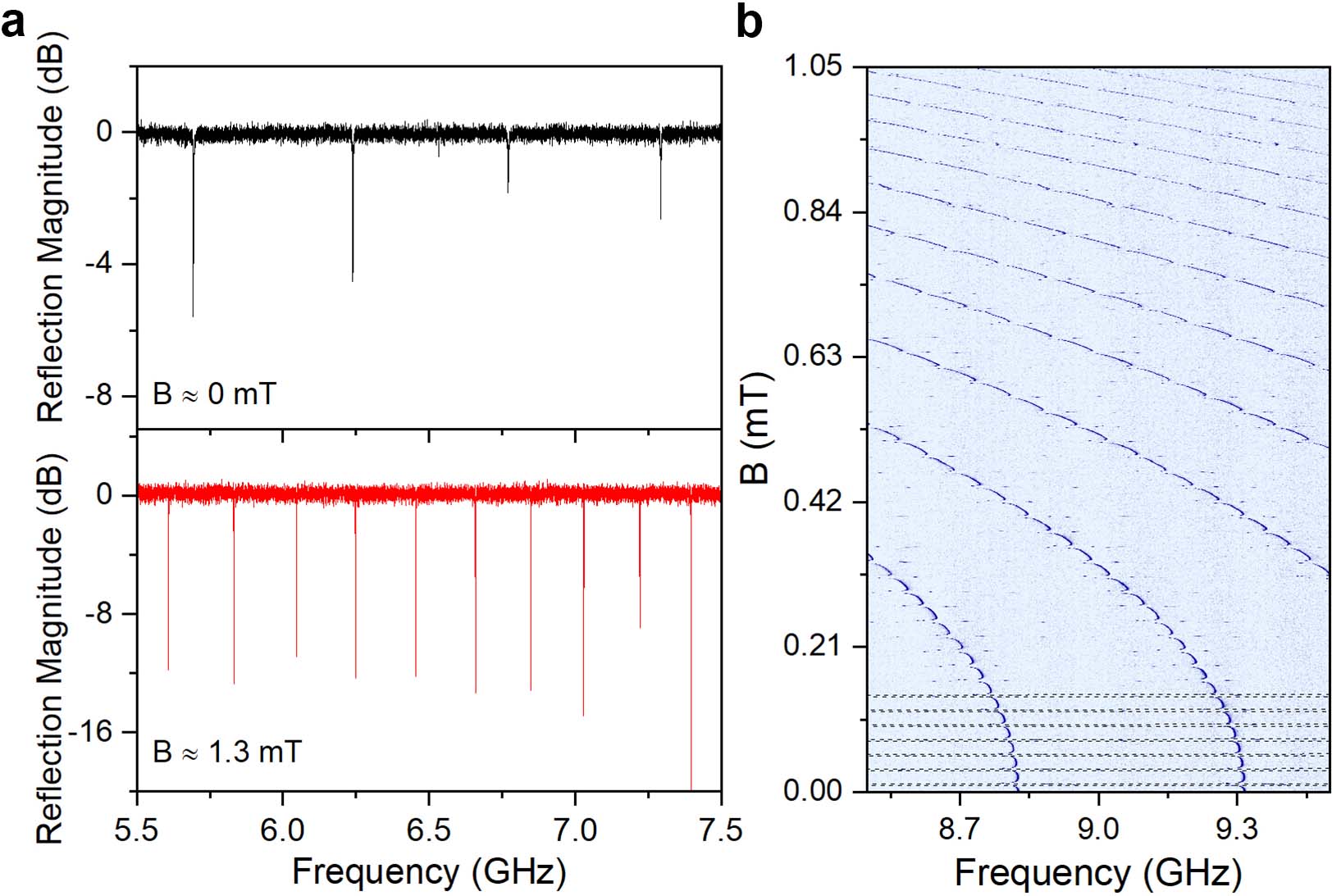}
	
	\caption{(a) The reflection magnitude as a function of the drive frequency at two values of the transverse magnetic field: $B\approx0~\textrm{mT}$ (top) and $B\approx1.3~\textrm{mT}$ (bottom). At both fields the qubit's resonance is maximally detuned, and it does not perturb the standing wave resonances of the transmission line. (b) The reflection magnitude as a function of the drive frequency and the transverse magnetic field. The color contrast is optimized to make the resonance maximally visible. The black dashed line shows the trajectory of the qubit's resonance for its first seven periods.}
	
	\label{fig:Fig2}
\end{figure}

Our test device is a Josephson transmission line made in coplanar stripline geometry with two parallel arrays of 3300 Josephson junctions (Fig. 1c). Fabricated using the standard Dolan bridge technique, the junctions are $3~\mu\textrm{m}$ wide, $0.4~\mu\textrm{m}$ long and separated by $0.2~\mu\textrm{m}$. Defining $L$ as twice the single junction's inductance and $C$ as the capacitance between two arrays per $0.6~\mu\textrm{m}$, our two arrays are equivalent to the telegraph transmission line in Fig. 1b. On the left, the arrays are attached to a dipole antenna for spectroscopy. On the right, the arrays are terminated with a single split Josephson junction. Similarly to ref. \cite{Kuzmin2021}, the split-junction behaves as a galvanically shunted transmon, which we use here as a probe of the transmission line impedance. We also use the flux periodicity of the split-junction to calibrate the magnetic flux magnitude around the device. The magnetic flux is created with a handmade superconducting coil.


\begin{figure}[b]
	\centering
	\includegraphics[width=1\linewidth]{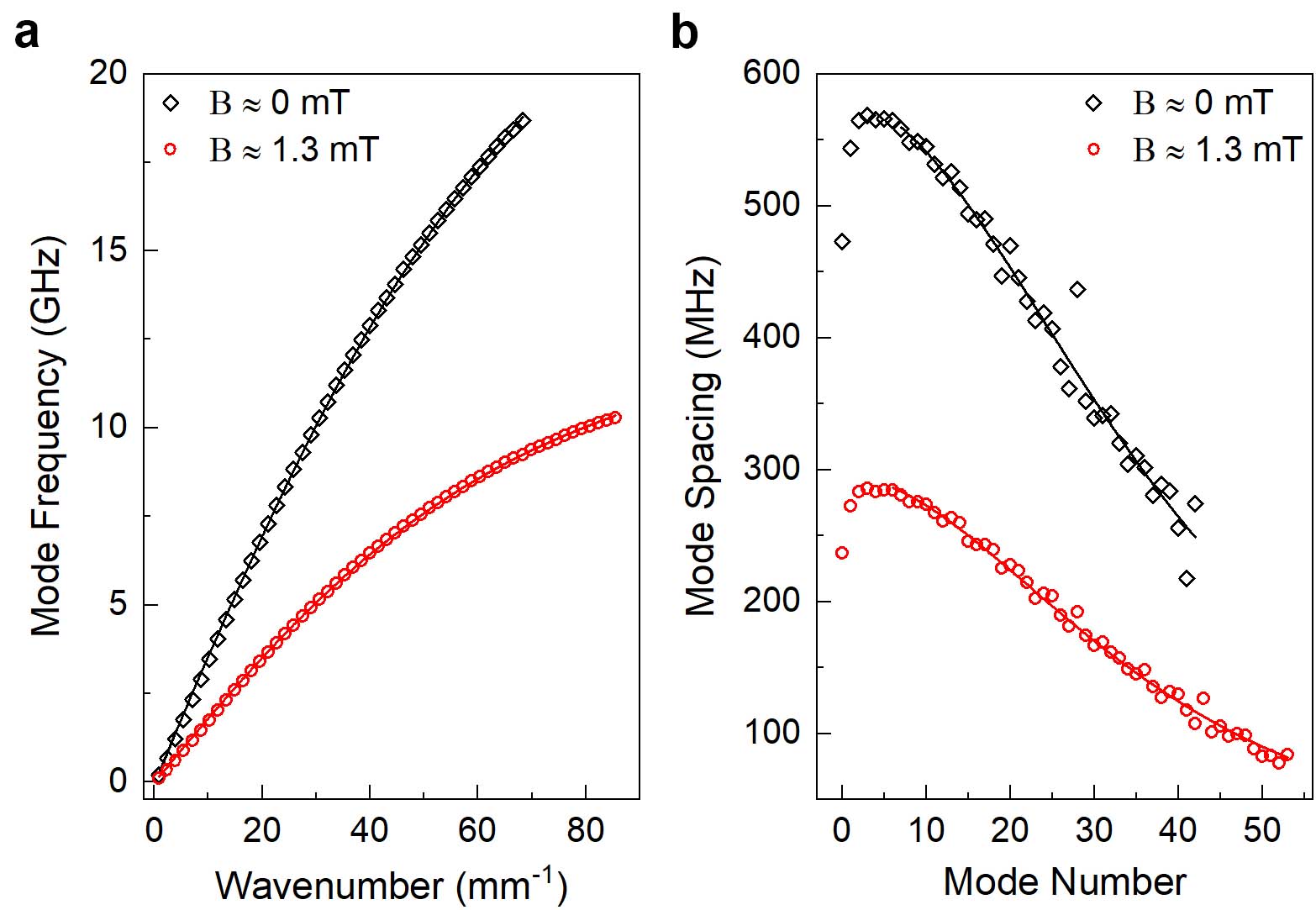}
	\caption{(a) The dispersion relation (markers) and its theoretical fit (lines) at two values of the transverse magnetic field. (b) The mode spacing (markers) as a function of the mode index and its theoretical fit (lines) at the same values of the transverse magnetic field.}
	
	\label{fig:Fig3}
\end{figure}

Following the well-established technique \cite{Kuzmin2019sit}, we performed the RF spectroscopy of our device in a varying transverse magnetic field. A typical spectrum reveals a set of almost equally spaced resonances corresponding to the standing wave modes of the Josephson transmission line (Fig. 2a). As the transverse field increases, the modes smoothly move to lower frequencies becoming more and more dense (Fig. 2b). The periodic in the field modulation, visible in Fig. 2b, is the result of the hybridization between the transmission line modes and the passing qubit's resonance. Its trajectory for the first seven periods is marked with a dashed black line. When the flux through the split-junction's loop $\Phi$ is equal to the integer number of flux quanta $\Phi_0$, the qubit's resonance is tuned above 20 GHz, where it does not perturb the transmission line's spectra. At the field $B\approx 1.3~\textrm{mT}$, the spacing between the resonances decreases by a factor of more than two, indicating a corresponding reduction in the wave velocity (Fig. 2a top and bottom). We note that from a separate measurement of the modes' lineshapes we did not notice any significant effect of the magnetic field on the internal loss in our Josephson junction arrays.


With the spectroscopy data, we reconstructed the wave's dispersion at two values of the transverse magnetic field (Fig. 3a). In the long-wavelength limit, the dispersion is linear, with the slope given by the wave velocity $v$. At the shorter wavelengths, the self-capacitance $C_J$, which shunts each Josephson junction, becomes important. This results in the saturation of the dispersion towards the plasma frequency $\omega_p=1/\sqrt{LC_J}$. Overall, the measured dispersion fits very well to a simple expression $2\pi f=vk/\sqrt{1+(vk/\omega_p)^2}$, where $k$ is a wavenumber. The dispersion provides us with the values of $v$ and $\omega_p$ which we used to find the wave impedance $Z$, following the procedure in ref. \cite{Kuzmin2019sit} (see Table \ref{tab:table1}).

Looking at the numbers, we see that a small $\sim1~\textrm{mT}$ transverse magnetic field tunes the wave velocity and the impedance by a factor of two, which corresponds to an increase in the arrays' inductance by a factor of four. In fact, the increase would be even greater at slightly higher fields, but the modes' decoupling from the antenna obstructs our spectroscopy. Such a strong effect of the transverse field on the array's inductance is the result of the proper geometry of our Josephson junctions. In particular, the junction's width to the length ratio needs to be much greater than one, which is $7.5$ in our case. We checked that making this number twice smaller significantly suppresses the demagnetization effect at similar transverse magnetic fields. This agrees with the observations and the theoretical predictions for a single Josephson junction \cite{Monaco2008,Monaco2009}.

Note that the transverse magnetic field does not add any disorder into the junction arrays. Figure 3b shows the measured spacings between the consecutive modes and their average given by the wave's dispersion. The fluctuations in the mode spacings around their average is a signature of a small fabrication disorder in the arrays' junctions. Importantly, the fluctuations do not grow at the higher field. This means that the transverse magnetic field tunes the arrays' inductance uniformly along the arrays' length.

\begin{figure}[t]
	\centering
	\includegraphics[width=1\linewidth]{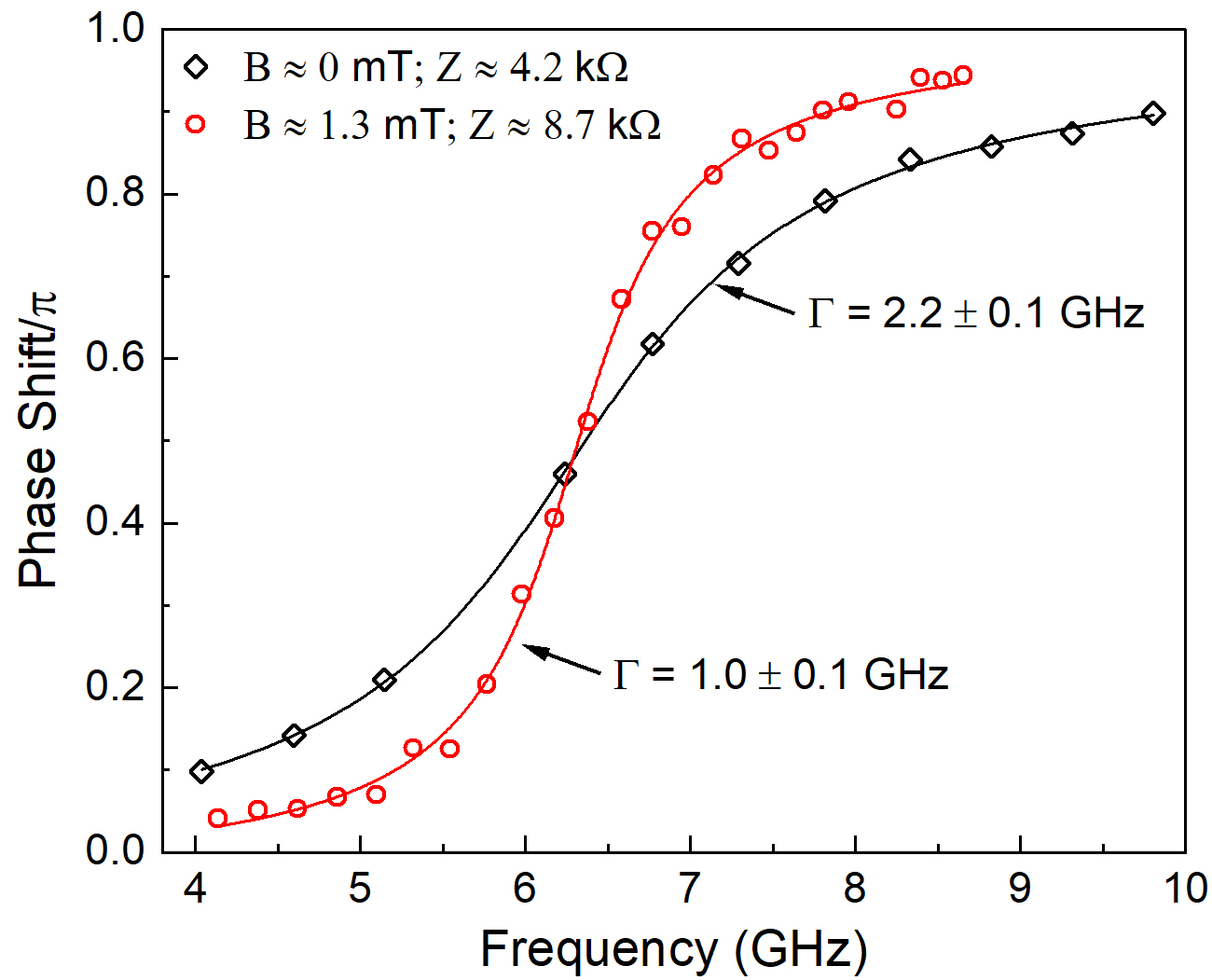}
	\caption{The frequency dependence of the phase shift introduced by the qubit at $f_0\approx6.1~\textrm{GHz}$ to the standing wave modes at two values of the transverse magnetic field. The phase shift is extracted from the qubit-induced shift in the modes positions, following the procedure described in ref \cite{Kuzmin2021}. The solid lines are the result of the theoretical fits.}
	
	\label{fig:Fig4}
\end{figure}


Finally, we use the qubit's resonance to confirm the change in the transmission line impedance. As was shown in ref. \cite{Kuzmin2019npj}, the resonance linewidth $\Gamma$ of a galvanically shunted transmon is inversely proportional to the transmission line's impedance $\Gamma=4R_QE_C/\pi Z$, where $E_C\approx1.1~\textrm{GHz}$ is the split-junction's charging energy. Tuning the qubit's resonance to the frequency $f_0\approx6.1~\textrm{GHz}$, we measured its linewidth at two values of the transverse magnetic field (Fig. 4). The quantity plotted in Fig. 4 is the phase shift $\delta(f)$, which the waves acquire at the qubit boundary and which was extracted through the modes frequency shifts \cite{Kuzmin2021}. As is expected for a Lorentzian resonance, the phase shift is well described by the expression $\delta(f)=\pi\arctan((f-f_0)/\Gamma)$ (solid lines in Fig. 4), providing the value of $\Gamma$. The measured values of the qubit's linewidth confirm the twofold increase in $Z$ at $B\approx1.3~\textrm{mT}$.


In summary, an array of single overlap-type Josephson junctions with a proper geometry allows for the tuning of its inductance by a transverse magnetic field. Using the junctions with the width to length ratio $7.5$, we demonstrated a fourfold change in the array's inductance at the field, comparable to those typically required to tune SQUID arrays. Reaching stronger tunability will require junctions with a larger width to length ratio. Our observations bring flux-tunability to single Josephson junction arrays, simplifying many experiments and extending the capabilities of Josephson junction based metamaterials.

The range of impedances demonstrated here makes our Josephson transmission line an ideally suited for analog quantum simulation of strongly interacting quantum impurity problems with a special point at $Z=R_Q$ \cite{gogolin2004bosonization}. By keeping the aspect ratio of the junctions the same, but changing their area (or unit cell size), it should be possible to move the range of tunability toward impedance values around $R_Q/2$ or $2R_Q$, covering other special points. In particular, our setup could bring clarity to the problem of the dissipative quantum phase transition, the existence of which, despite many previous attempts, stays controversial \cite{Murani2020}.

\begin{table}[htbp]
  \centering
  \caption{The device's parameters at two values of the transverse magnetic field $B$. The errors for the wave velocity $v$, plasma frequency $\omega_p/2\pi$, and qubit's resonance linewidth $\Gamma$ are the standard errors of the fits. For the wave impedance $Z$, the error is half of the difference between impedance values calculated using two approaches detailed in \cite{Kuzmin2019sit}.}
    \begin{tabular}{|r|l|l|l|l|}
    \toprule[0.5pt]
    \multicolumn{1}{|l|}{$B$, $\textrm{mT}$} & $v$, $10^6 \textrm{m}/\textrm{sec}$ & $\omega_p/2\pi$, GHz & $Z$, $\textrm{k}\Omega$ & $\Gamma$, GHz \\
    \midrule[0.5pt]
    0     & 2.33$\pm$0.03 & 28.7$\pm$0.3 & 4.2$\pm$0.1 & 2.2$\pm$0.1 \\
    \midrule[0.5pt]
    1.3   & 1.19$\pm$0.01 & 13.7$\pm$0.1 & 8.7$\pm$0.6 & 1.0$\pm$0.1 \\
    \bottomrule[0.5pt]
    \end{tabular}%
  \label{tab:table1}%
\end{table}%

\textbf{Acknowledgements.} 
This work was funded by US DOE through the Early Career Award No. DE-SC0020160; US-Israel Binational Science Foundation through Grant No. 2020072.

\bibliography{TunableImpedanceReferences}
\end{document}